\documentclass[12pt,preprint,useAMS,usenatbib]{emulateapj}
\usepackage{graphicx}
\usepackage{subfloat}
\usepackage{amsmath}
\usepackage{url}
\usepackage{color,ulem}
\usepackage[hypertex]{hyperref}

\def\msun{\,{M_\odot}}

\def\sfr{{M_\odot\,{\rm yr^{-1}}}}

\def\spose#1{\hbox to 0pt{#1\hss}}
\def\lta{\mathrel{\spose{\lower 3pt\hbox{$\mathchar"218$}}
     \raise 2.0pt\hbox{$\mathchar"13C$}}}
\def\gta{\mathrel{\spose{\lower 3pt\hbox{$\mathchar"218$}}
     \raise 2.0pt\hbox{$\mathchar"13E$}}}

\newcommand{\etal}{{et al.}}
\def\kms{\,{\rm km\,s^{-1}}}
\def\kmsmpc{\,{\rm km\,s^{-1}\,Mpc^{-1}}}

\def\cm3{\,{\rm cm^{-3}}}

\lefthead{Madau, Haardt, \& Dotti}
\submitted{Received 2014 February 26; accepted 2014 March 4}


\begin{document}

\title{Super-Critical Growth of Massive Black Holes From Stellar-Mass Seeds}

\author{Piero Madau$^{1}$, Francesco Haardt$^{2,3}$, and Massimo Dotti$^{3,4}$}
\altaffiltext{1}{Department of Astronomy and Astrophysics, University of California, 1156 High Street, Santa Cruz, CA 95064, USA.}
\altaffiltext{2}{Dipartimento di Scienza e Alta Tecnologia, Universit\`a dell'Insubria, via Valleggio 11, I-22100 Como, Italy.}
\altaffiltext{3}{INFN, Sezione di Milano-Bicocca, Piazza della Scienza 3, I-20126 Milano, Italy.}
\altaffiltext{4}{Dipartimento di Fisica G. Occhialini, Universit\`a degli Studi di Milano Bicocca, Piazza della Scienza 3, I-20126 Milano, Italy.}

\begin{abstract}
We consider super-critical accretion with angular momentum onto stellar-mass black holes as a possible mechanism for growing
billion-solar-mass holes from light seeds at early times. We use the radiatively-inefficient ``slim disk" solution -- 
advective, optically thick flows that generalize the standard geometrically thin disk model -- to show how mildly 
super-Eddington intermittent accretion may significantly ease the problem of assembling the first massive black holes when 
the universe was less than 0.8 Gyr old. Because of the low radiative efficiencies of slim disks around non-rotating as well 
as rapidly rotating holes, the mass $e$-folding timescale in this regime is nearly independent of the spin parameter. The 
conditions that may lead to super-critical growth in the early universe are briefly discussed. 
\end{abstract}


\section{Introduction} \label{intro}

The most distant quasar discovered to date, ULAS J1120+0641 at a redshift $z=7.084$, is believed to host a black hole with a 
mass of $2.0^{+1.5}_{-0.7}\times 10^9\,\msun$ that is shining 0.78 Gyr after the big bAng \citep{Mortlock11}. This object, together with the handful 
of bright {Sloan Digital Sky Survey} (SDSS) quasars at redshift $z\gta 6$ \citep{Fan06}, 
sets some of the tightest constraints on models for the formation and growth of massive black holes (MBHs) at early epochs.   
The challenge provided by the existence of billion-solar-mass black holes at the end of the reionization epoch is easily described (see, e.g., \citealt{Haiman13}
for a review). If MBHs are assembled by the accretion of gas onto less massive ``seed" holes at the rate $\dot m$, and if in the process a fraction $\epsilon$ 
of the rest-mass energy of the infalling material is released as radiation, then the growth of the hole's mass $M$ is regulated by the equation
\begin{equation}
{dM\over dt}=(1-\epsilon)\dot m=\left({1-\epsilon\over \epsilon}\right)\left({L\over L_E}\right){M\over t_E}.
\end{equation}
Here $L$ is the radiated luminosity, $\epsilon \equiv L/\dot m c^2$, $L_E \equiv 4\pi G M \mu_e m_pc/\sigma_T$ is the Eddington limit when
the continuum radiation force balances gravity, $\sigma_T$ is the Thomson scattering cross-section, $\mu_e$ is the mean molecular weight 
per electron, $t_E\equiv Mc^2/L_E=0.44\mu_e^{-1}$ Gyr is the Eddington timescale, and all other symbols have their usual meaning.    
The characteristic $e$-folding timescale $t_{\rm acc}$ for mass growth is then 
\begin{equation}
t_{\rm acc}=\left({\epsilon\over 1-\epsilon}\right)\left({L_E\over L}\right)t_E=(4.3\times 10^7~{\rm yr})\,\left({L_E\over L}\right), 
\label{eq:tacc}
\end{equation}
where the last equality assumes $\mu_e=1.15$ (valid for primordial gas) and a radiative efficiency of $\epsilon=0.1$.  In a concordance 
cosmology with $\Omega_M=0.27$, $\Omega_\Lambda=0.73$, and $H_0=70\,\kmsmpc$, the time elapsed between 
$z=20$ and $z=7$ is 0.6 Gyr, corresponding to 14 $e$-foldings of Eddington-limited accretion ($L=L_E$) and a mass amplification factor of $10^{6}$. 

The growth of $2\times 10^9\,\msun$ MBHs at the Eddington rate from light black hole seeds of mass $M_0=100\,\msun$ requires $\ln(2\times 10^9/100)=17$ 
$e$-foldings at $L=L_E$. Therefore, if the first seeds were $\sim 100\,\msun$ remnants of the first generation of massive stars 
\citep[e.g.,][]{Madau01,Haiman01,Heger03,Volonteri03}, these could grow into billion-solar-mass holes by $z\sim 7$ only if all the following 
conditions were fulfilled: (1) seeds were present early on, at $z\gta 20$; (2) gas accretion continued more or less uninterrupted at the Eddington 
rate for $\gta 0.6$ Gyr; and (3) $\epsilon<0.1$ \citep{Tanaka09}. The second condition is hard to satisfy in the shallow potential wells of low-mass dark 
matter halos, as feedback effects resulting from the accretion process itself are expected to dramatically affect gas inflow and may result in sub-Eddington 
rates and negligible mass growth \citep{Johnson07,Pelupessy07,Alvarez09,Milos09}. The third condition requires radiative efficiencies that are below 
those expected for thin disk accretion onto rapidly spinning Kerr black holes ($\epsilon \simeq 0.3-0.4$, \citealt{Thorne74,Shapiro05}), and approach the 
value, $\epsilon=0.057$, characteristic of the Schwarzschild non-rotating solution.

Over the last decade, a number of alternatives to the above picture have been proposed. If stellar seeds were present in large numbers at high redshifts, 
coalescing black hole binaries brought together by successive galaxy mergers may, in principle, help mass build-up and generate mass amplification
factors as high as $10^4$ \citep{Yoo04}. More massive seeds, with $M_0\sim 10^4-10^5\,\msun$, may form through the ``direct collapse" of low angular 
momentum gas at high redshift \citep[e.g.,][]{Loeb94,Bromm03,Koushiappas04,Lodato06,Spaans06,Regan09,Mayer10}, likely via the intermediate stage of supermassive stars 
\citep{Begelman10}, and therefore ``jump start" the whole process.  Questions remain about the idealized conditions needed in these models to avoid 
fragmentation, dissipate angular momentum, and drive gas towards the center of protogalaxies at extremely high rates.

In this {Letter} we discuss {\it super-critical (i.e., super-Eddington) accretion with angular momentum onto stellar-mass seeds} as a possible mechanism 
for bypassing some of the above difficulties. Evidence for near-Eddington or super-Eddington flows has been accumulating in recent years. Super-critical 
accretion onto stellar-mass black holes has been invoked to explain the nature of the ultraluminous X-ray sources \citep[e.g.,][]{Gladstone09,Middleton13}. 
A study of a large sample of active galactic nuclei (AGNs) suggests that 
many of them emit considerably more energy and have higher $L/L_E$ ratios than previously assumed \citep{Netzer14}. \citet{Kormendy13} have recently argued
that the normalization of the local black hole scaling relations should be increased by a factor of five to $M=0.5\%\,M_{\rm bulge}$. 
This increases the local mass density in black holes by the same factor, decreases the required mean radiative efficiency to 1\%-2\%,  
and may be evidence for radiatively-inefficient super-Eddington accretion \citep[e.g.,][]{Soltan82,Novak13}. At high redshifts, the 
very soft X-ray spectrum of ULAS J1120+0641 appears to suggest that this quasar 
is accreting at super-critical rates \citep{Page13}. On the theoretical side, it is known that the standard, radiatively-efficient thin disk 
solution \citep{Shakura76} can no longer be applied when mass is supplied to a black hole at high rates. In this regime, viscosity-generated 
heat does not have sufficient time to be radiated away, and is instead advected into the hole. The shorter mass $e$-folding timescales and the 
decreased radiative efficiencies that characterize these flows make them ideal for feeding and growing MBHs out of stellar-mass seeds at early times.    

\section{Super-critical accretion: the slim disk solution} \label{slim}

The Shakura-Sunyaev treatment of accretion onto a black hole via a thin disk posits a radiatively efficient flow where all the heat generated by 
viscosity at a given radius is immediately radiated away. It is a {\it local} model, described by algebraic equations, valid at any particular 
(radial) location in the disk, independently of the physical conditions at different radii. At high accretion rates, i.e., when $\dot m\gta 0.3\,\dot m_E$, 
this assumption is incorrect. Here, $\dot m_E\equiv 16 L_E/c^2$ is the critical accretion rate that gives origin to about an Eddington luminosity in the 
case of a radiatively-efficient thin disk around a non-rotating black hole.\footnote{Note that many authors use a different definition of the critical accretion 
rate, i.e., $\dot m_E\equiv L_E/c^2$.}\,Optically thick, stationary ``slim disks" offer a more general set of {\it non-local} solutions,
obtained by numerical integration of the two-dimensional stationary Navier-Stokes equations with a critical point - the radius at which the 
gas velocity exceeds the local speed of sound \citep{Abramowicz88}. 

\begin{figure*}[ht]
\centering
\includegraphics[width=0.49\textwidth]{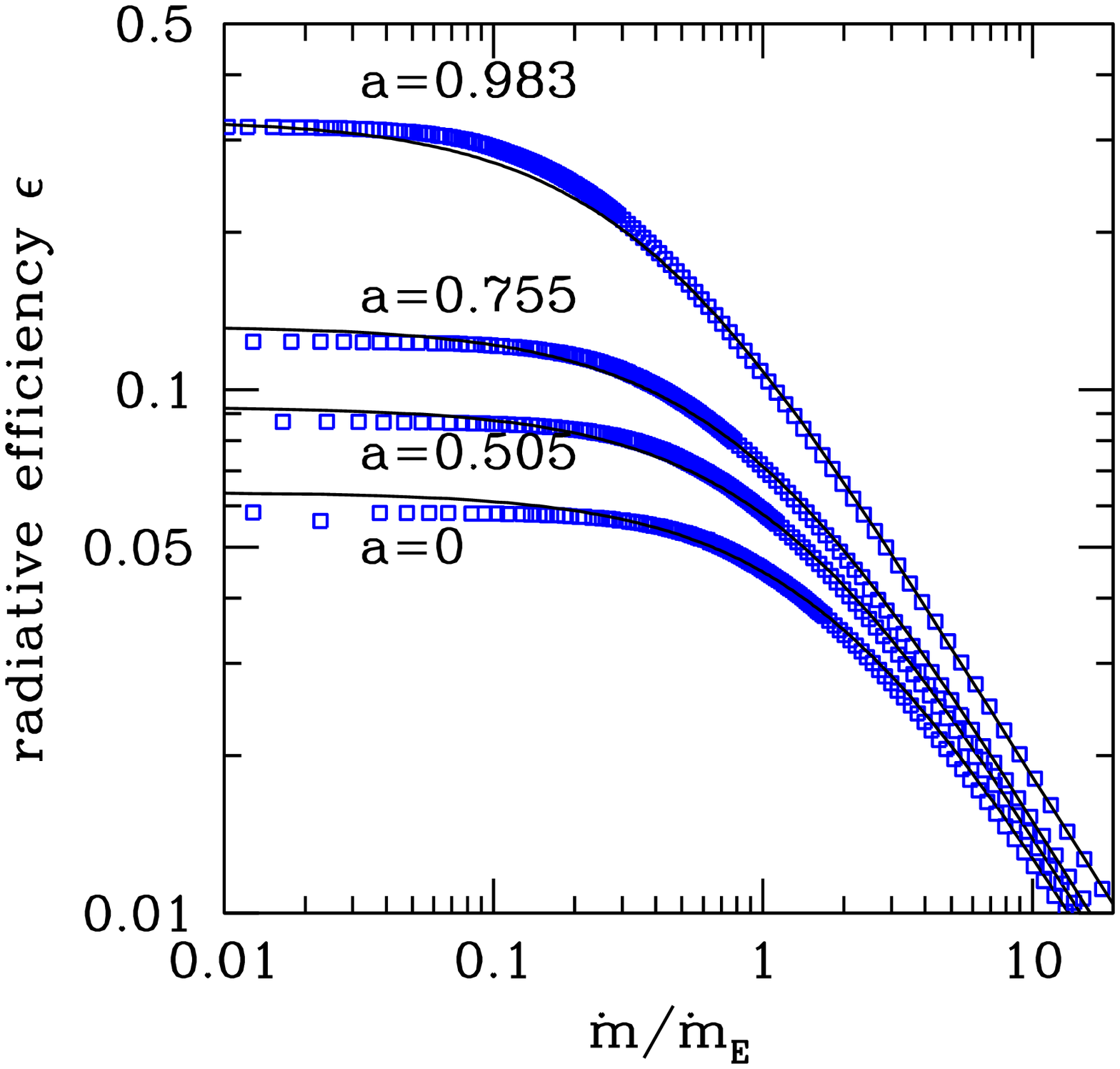}
\includegraphics[width=0.49\textwidth]{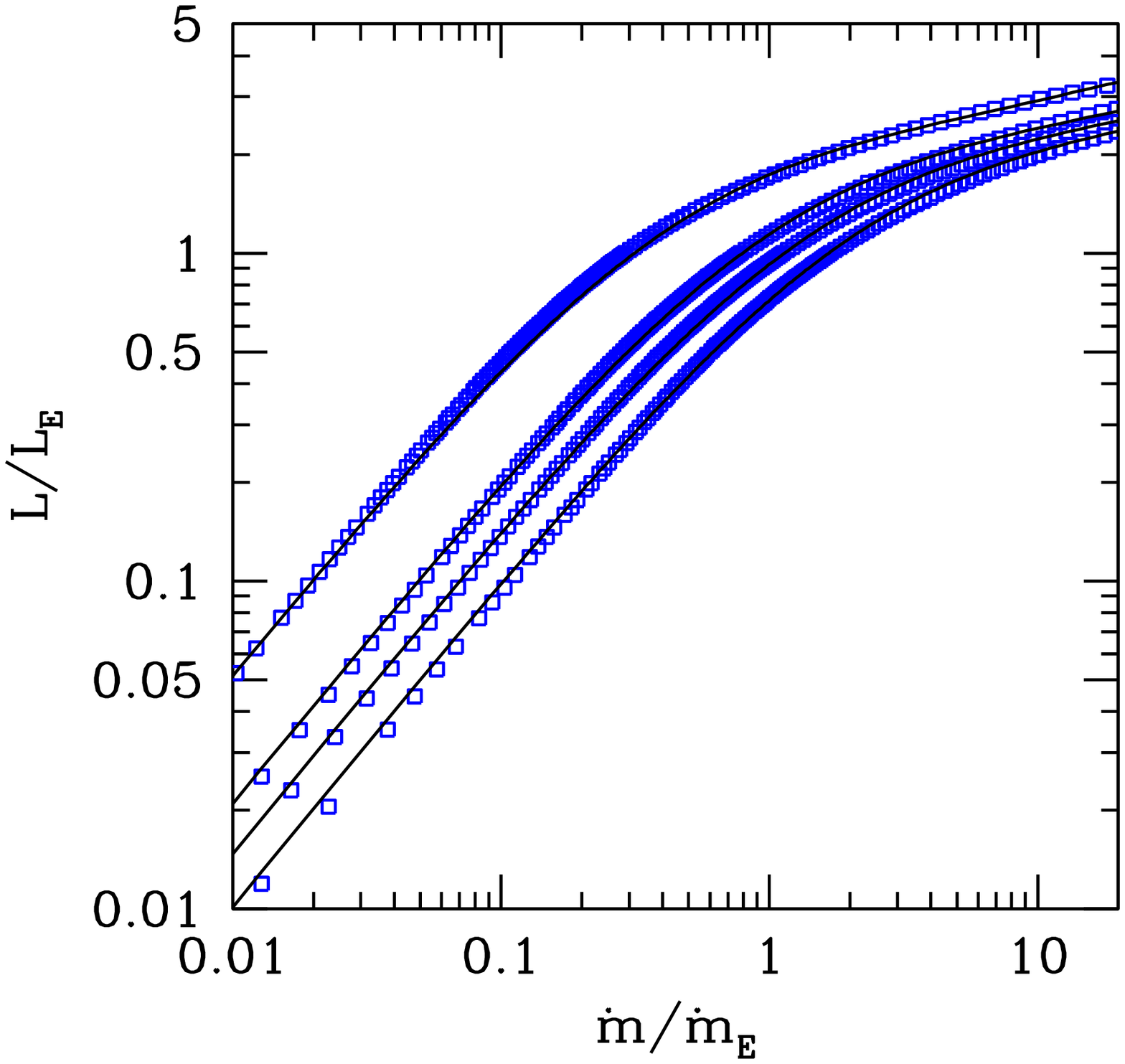}
\caption{Radiative efficiency $\epsilon=L/{\dot m}c^2$ and total luminosity $L/L_E$ of an accreting black hole are plotted in the left and right panels 
as a function of the accretion rate $\dot m$ (in units of the Eddington rate $\dot m_E\equiv 16L_E/c^2$). The blue points are the results 
of the numerical integration of the relativistic slim disk equations obtained by \citet{Sadowski09}, while the solid curves from top-to-bottom 
show our best-fit function (Equation (\ref{eq:ellfit})) for different spin parameters. 
}
\label{fig1}
\end{figure*}

\begin{figure*}[ht]
\centering
\includegraphics[width=0.49\textwidth]{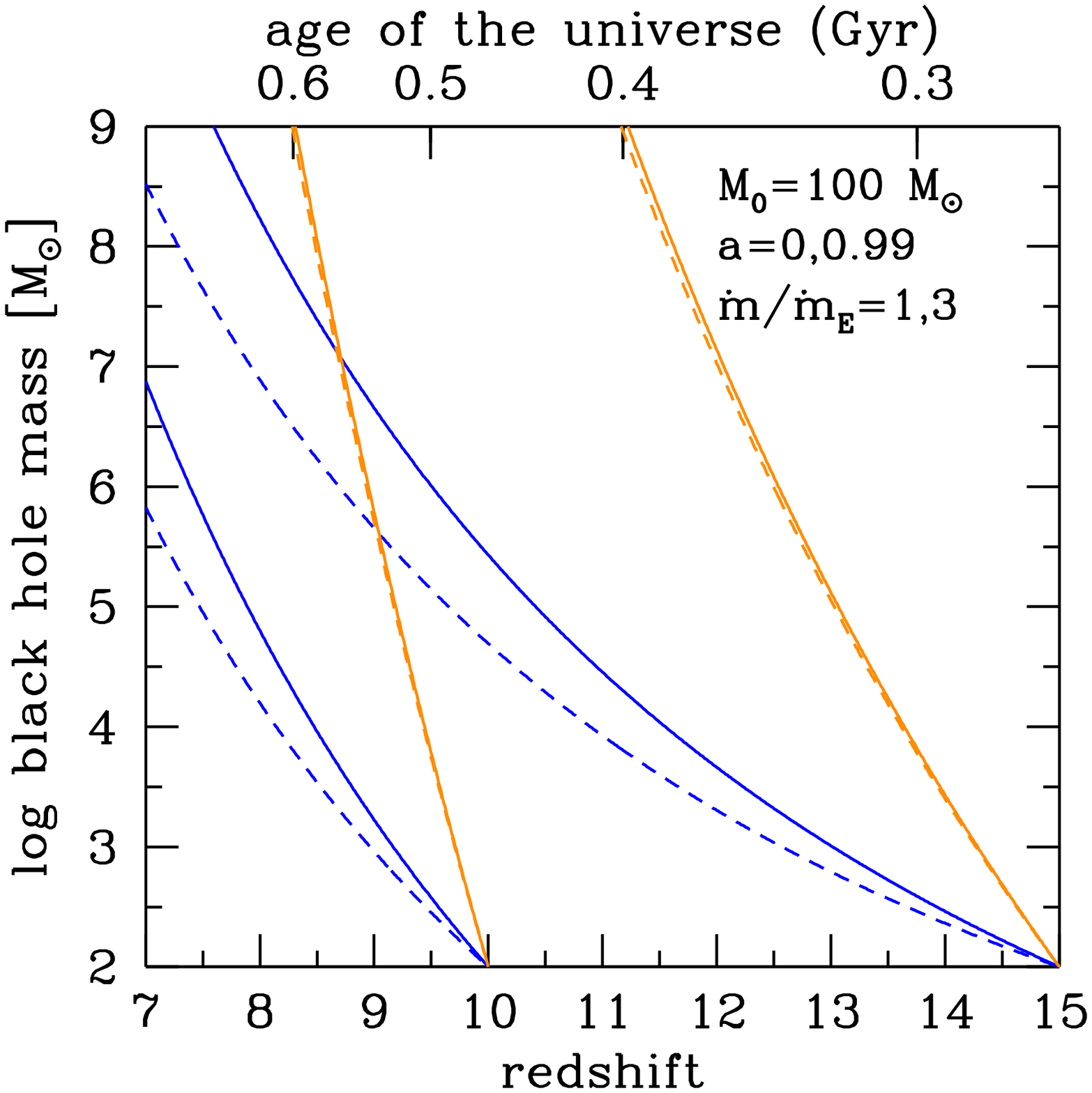}
\includegraphics[width=0.49\textwidth]{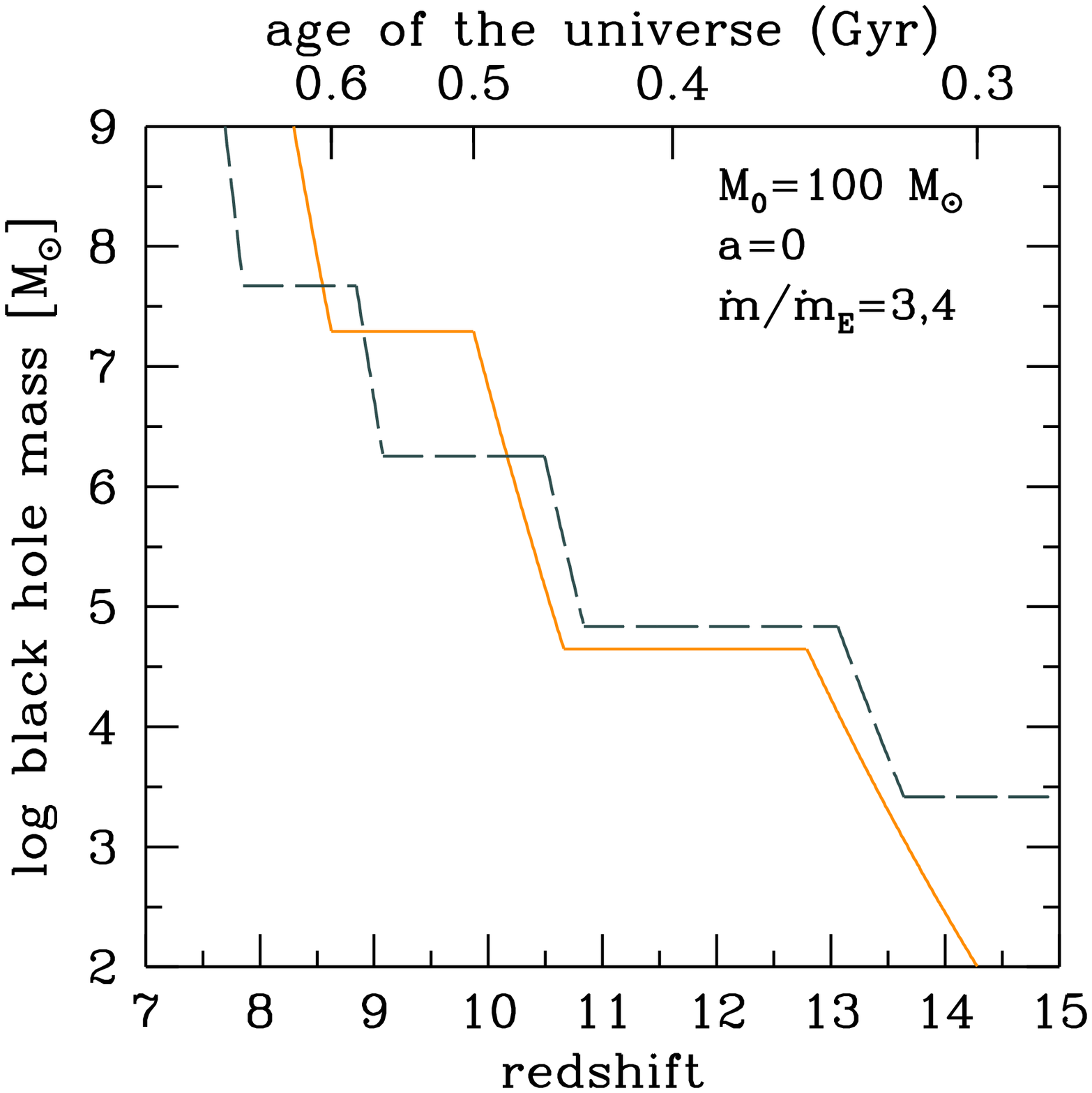}
\caption{{Left panel:} mass assembly history of a seed black hole with $M_0=100\,\msun$ that is accreting continuously from redshifts 10 and 15. 
{\it Orange curves}: slim-disk accretion at $\dot m/\dot m_E=3$ for (constant) spin parameter $a=0$ ({\it solid lines}) and $a=0.99$ ({\it 
dashed lines}).  {\it Blue curves}: same but for Eddington-limited ($\dot m/\dot m_E=1$) growth according to the classical thin disk solution.
{Right panel:} mass assembly history of a seed, non-rotating hole with $M_0=100\,\msun$ accreting intermittently at $\dot m/\dot m_E=3$ 
with a duty cycle of 0.5 ({\it orange solid line}) and at $\dot m/\dot m_E=4$ with a duty cycle of 0.2 ({\it gray dashed line}). 
In both cases the period of quiescence is 100 Myr.  
}
\label{fig2}
\end{figure*}

To model such super-critical flows -- which are characterized by large radial velocities, non-Keplerian rotation, inner edges that are closer to the 
black hole than the innermost stable circular orbit, and low radiative efficiencies -- we use the numerical solutions of the relativistic slim 
accretion disk equations obtained by \citet{Sadowski09} and available online at 
\url{http://users.camk.edu.pl/as/slimdisk.html}. Figure \ref{fig1} shows how super-critical accretion is qualitatively different from the standard, 
sub-Eddington, thin disk solution. The right panel depicts the disk luminosity (in units of $L_E$) versus the accretion rate 
(in units of $\dot m_E$) for four values of the black hole spin parameter $a=0.983, 0.755, 0.505,$ and $0$. The corresponding radiative efficiency 
$\epsilon=L/{\dot m}c^2$ is plotted in the left panel. Despite super-Eddington $\dot m>\dot m_E$ accretion rates, 
slim disks remain only moderately luminous ($L\gta L_E$), as a large fraction of the viscosity-generated heat is advected inward and released 
closer to the hole or not released at all. As a result of the increasing rate of advection, the efficiency of transforming gravitational energy 
into radiative flux decreases with increasing accretion rate. For ease of use and flexibility, we have fitted the two dimensional tabulated 
luminosity as 
\begin{equation}
L/L_E=A(a) \left[ \frac{0.985}{\dot m_E/\dot m+B(a)}\,+\,\frac{0.015}{\dot m_E/\dot m+C(a)}\right],
\label{eq:ellfit}
\end{equation}
where the functions $A, B,$ and $C$ scale with the spin of the black hole as
\begin{eqnarray}
A(a) & = & (0.9663-0.9292a)^{-0.5639},\\
B(a) & = & (4.627-4.445a)^{-0.5524},\\
C(a) & = & (827.3-718.1a)^{-0.7060}.
\end{eqnarray}
Our fits to the emitted luminosity and ensuing radiation efficiency are compared in Figure \ref{fig1} to the numerical results of \citet{Sadowski09}.
Over the range $0.001<\dot m/\dot m_E<500$ and $0<a<0.998$, fit residuals are typically below 7\%.   
Since, in the case of photon-trapped supercritical accretion, the emitted luminosity is not linearly proportional to the accretion rate,     
it is convenient to rewrite Equation (\ref{eq:tacc}) as 
\begin{equation}
t_{\rm acc}={t_E\over 16(1-\epsilon)}\left({\dot m_E\over \dot m}\right)\lta (8.4\times 10^6~{\rm yr})\,\left({3\dot m_E\over \dot m}\right), 
\label{eq:tsacc}
\end{equation}
where the last inequality holds for modestly super-Eddington rates independently of the value of the black hole spin. Figure \ref{fig2} shows how even small modifications 
to accretion rates and radiative efficiencies can have an exponential impact on the growth of seed black holes. In the left panel the cosmic assembly 
history of a seed hole of initial mass $M_0=100\,\msun$ accreting at $\dot m/\dot m_E=3$ from redshifts 10 and 15 is compared to an 
Eddington-limited ($\dot m/\dot m_E=1$) growth that follows the classical thin disk solution. 
Two curves are shown, for (constant) spin parameter $a=0$ and $a=0.99$. Because of the low radiative efficiencies
of slim disks around non-rotating as well as rapidly rotating holes, the mass $e$-folding timescale in this regime is nearly independent of the
spin parameter. This is in contrast to the thin disk solution, where the mass of the growing hole is exponentially sensitive to its spin.   

From the astrophysical standpoint, however, it seems unlikely that early-growing black holes may be able to sustain uninterrupted 
super-critical accretion rates for half a Gyr or so. The right panel of Figure \ref{fig2} shows the illustrative growth histories of: (1) a 
seed non-rotating hole undergoing three major episodes of $\dot m/\dot m_E=3$ accretion each lasting 50 Myr followed by a 100 Myr 
period of quiescence, i.e., a duty cycle of 0.5; and (2) a seed non-rotating hole undergoing five major episodes of $\dot m/\dot m_E=4$ 
accretion each lasting 20 Myr followed by a 100 Myr period of quiescence, i.e., a duty cycle of 0.2. We have chosen a 100 Myr quiescence period
since this is the mean time interval between major mergers (mass ratios $\ge$ 1:3) at redshift 14 for a $10^{10}\,\msun$ descendant 
halo \citep{Fakhouri10}. High duty cycles of 0.2-0.5 match those inferred from the observed clustering strength of bright quasars at redshift $z=3-4.5$
\citep{Shankar10}. Clearly, the shorter mass $e$-folding timescales
of flows that are only mildly super-critical can significantly ease the problem of assembling MBHs out of stellar-mass seeds at early times even in the case
of intermittent accretion. We note here that, while even shorter duty cycles may lead to the growth of MBHs if accretion was occurring at highly 
super-Eddington rates, $\dot m/\dot m_E\gg 10$, the slim disk solution is not directly applicable in such regimes. Indeed, general relativistic 
magnetohydrodynamic simulations of black hole accretion at rates in the range $\dot m/\dot m_E=20-200$ have recently shown that these flows are actually  
efficient in terms of the total energy escaping from the system, which is mostly in the form of thermal and kinetic energy of outflowing gas 
and Poynting flux \citep{Sadowski14,McKinney13}. In highly super-critical flows, the magnitude of the outflow is found to be comparable with the   
inflow accretion rate \citep{Sadowski14}.

\section{Discussion} 
%\section{Super-critical growth in the early universe} \label{highz}

The assembly of pregalactic MBHs following a phase of super-critical quasi-spherical accretion in metal-free halos has been 
explored by \citet{VolonteriRees05}. Highly super-critical rates at early times have been recently advocated by \citet{Volonteri14}.   
Here, we have extended and updated these works by focusing on the ``slim disk" solution --  advective, optically thick 
disks that generalize the standard model of radiatively efficient thin flows to moderately super-Eddington accretion rates. 
Under the assumption that a gas supply rate of a few $\times\, \dot m_E\approx 0.01\,M_5\,\sfr$ (where $M_5$ is the mass of the hole in units of $10^5\,\msun$) 
is indeed able to reach the central MBH for a period of 20 Myr or so, we have shown how a few episodes of super-critical accretion may turn a 
handful of light seed holes into the population of rare bright quasars observed at $z\gta 6$ by the SDSS. 
Such high rates may both be determined by local physics within the accretion flow as well as by the large-scale cosmological environment 
in which MBHs and their hosts are growing. An example of small-scale physics is the presence of density inhomogeneities induced by 
radiative-hydrodynamic instabilities that reduce the effective opacity and allow super-Eddington fluxes from slim, porous disks \citep{Dotan11}.
On large scales, gas-rich major mergers between massive protogalaxies have been shown to drive large gas flows towards the center of the    
merger remnant, owing to continuous losses of angular momentum by torques and shocks over a wide range of spatial scale \citep[e.g.,][]{Mayer10}. 
Contrary to the formation of MBH seeds by rapid ``direct collapse", this gas supply is not limited by the Kelvin-Helmholtz contraction timescale 
of the central supermassive proto-star \citep[see, e.g.,][]{Shang10}, but can add to the central black hole mass 
over the longer accretion timescale of Equation (\ref{eq:tsacc}). And while these high-accretion episodes may generate AGN outbursts, drive hot, high-velocity 
outflows from the host galaxy, and occasionally clear the nuclear region of most of its gas, additional gas inflows may quickly rebuild
the nuclear reservoir as AGN feedback appears to couple only weakly to a galaxy disk \citep{Gabor14}.

On intermediate scales, models of circumnuclear disks supported by the turbulent pressure caused by supernova explosions predict high mass 
supply rates to the central 1 pc lasting for $10^8$ yr \citep{Kawakatu08}. Even in the absence of star formation, thick, turbulent 
pressure-supported disks are often seen in simulations of early protogalaxies \citep{Wise08}. Viscous torques generated by the dissipation of 
supersonic turbulent motions generate a mass accretions rate \citep{Pringle81} 
\begin{equation}
\dot m=2\pi \nu \Sigma_0 \, \vert \frac{d\ln \Omega}{d\ln r} \vert,
\label{eq:feeding}
\end{equation}
where $\Sigma_0$ is the central surface density of the disk, $\nu \simeq \sigma h$ is the viscous parameter \citep{Wada02}, $\sigma$ is the gas turbulent 
velocity, $h$ is the disk scale height, $h/r\sim \sigma/V_{\rm rot}$, and $V_{\rm rot}$ is the disk rotational velocity.
Observations show that, at a fixed stellar mass, galaxies systematically increase in disordered motions and decrease in rotation velocity with increasing 
redshift and decreasing potential \citep{Kassin12}. At $z\sim 2$, the ratio of the rotational to random velocities ranges between 1 and 6, quite 
in contrast to present-day disk galaxies where $V_{\rm rot}/\sigma\sim 10-20$ \citep{Forster09}. 

Assuming $V_{\rm rot}/\sigma\sim 1$ in the early universe,   
writing the disk mass as $M_d=2\pi \Sigma_0 R_d^2$ where $R_d$ is the disk scale length, and relating it to $V_{\rm rot}$ using the baryonic
Tully-Fisher relation, $M_d={\cal A}\,V_{\rm rot}^4$ \citep{McGaugh05}, Equation (\ref{eq:feeding}) yields an order-of-magnitude estimate 
of the accretion rate expected in this ``turbulent feeding" model,
%
%\begin{multline}
\begin{align}
\dot m \sim {\cal A} V_{\rm rot}^5 {r\over R_d^2} & \sim (1\,\sfr)\,\left({V_{\rm rot}/30\,\kms}\right)^5 \times \nonumber \\
& \times \left({r\over 0.1\,R_d}\right)\left({{\rm 0.1\,kpc}\over R_d}\right), 
\label{eq:supply}
%\end{multline}
\end{align}
where ${\cal A}=50\,\msun {\rm km^{-4}\,s^4}$ \citep{McGaugh05} and we have dropped numerical factors of order unity. 

The estimate above shows that large mass supply rates in the dense environments of high redshift, dispersion-dominated, massive protogalaxies are 
theoretically plausible. It is consistent with the mass accretion rates measured at 100 pc from the center of simulated atomic cooling halos at 
$z>10$ \citep{Prieto13}, and is comparable to the mass accretion rates, $\sim c_s^3/G\propto T^{3/2}/G$ expected in self-gravitating, 
collapsing $T\sim 10^4\,$K gas clouds. Rapid MBH growth may occur in rare special environments where gas can flow 
toward the center at super-Eddington rates relatively unaffected by feedback processes. Indeed it would seem fortuitous if the mass 
supply rate was always regulated exactly at $\dot m_E$ during the early growth of MBHs, as it is commonly assumed. Equally contrived, of course, 
are gas fueling rates that remain at $(3-4)\times \dot m_E$ for 20-50 Myr while the central MBHs are growing by many orders of magnitude in mass.
Perhaps a more plausible scenario is one where $\dot m\gg \dot m_E$ early on (when black holes are still of intermediate mass), but only a 
fraction of this external mass supply reaches the horizon, and where $\dot m$ ``dwindles" to a few times Eddington during the last few $e$-foldings, 
when the MBH is already above $10^8\,\msun$.

\acknowledgements
%\vspace{0.5cm}
Support for this work was provided by the NSF through grant OIA-1124453, and by NASA through grant NNX12AF87G (P.M.). 
We thank L. Mayer and the referee, Z. Haiman, for their constructive questions and comments on our manuscript.  

%\page


\begin{thebibliography}{}

\bibitem[Abramowicz \etal(1988)]{Abramowicz88} Abramowicz, M. A., Czerny, B., Lasota, J. P., \& Szuszkiewicz, E. 1988, ApJ, 332, 646

\bibitem[Alvarez \etal(2009)]{Alvarez09} Alvarez, M. A., Wise, J. H., \& Abel, T. 2009, ApJL, 701, L133

\bibitem[Begelman(2010)]{Begelman10} Begelman, M. C. 2010, MNRAS, 402, 673  

\bibitem[Bromm \& Loeb(2003)]{Bromm03} Bromm, V., \& Loeb, A. 2003, ApJ, 596, 34

\bibitem[Dotan \& Shaviv(2011)]{Dotan11} Dotan, C., \& Shaviv, N. J. 2011, MNRAS, 413, 1623
  
\bibitem[Fakhouri \etal(2010)]{Fakhouri10} Fakhouri, O., Ma, C.-P., \& Boylan-Kolchin, M. 2010, MNRAS, 406, 2267 

\bibitem[Fan(2006)]{Fan06} Fan, X. 2006, NewAR, 50, 665 

\bibitem[F\"orster Schreiber \etal(2009)]{Forster09} F\"orster Schreiber, N. M., Genzel, R., Bouche, N. 2009, ApJ, 706, 1364 

\bibitem[Gabor \& Bournaud(2014)]{Gabor14} Gabor, J. M., \& Bournaud, F. 2014, MNRAS, submitted (arXiv:1402.4482) 

\bibitem[Gladstone \etal(2009)]{Gladstone09} Gladstone, J. C., Roberts, T. P., \& Done, C. 2009, MNRAS, 397, 1836 

\bibitem[Haiman(2013)]{Haiman13} Haiman, Z. 2013, The First Galaxies (Astrophysics and Space Science Library, Volume 396; Berlin: Springer, 293

\bibitem[Haiman \& Loeb(2001)]{Haiman01} Haiman, Z., \& Loeb, A. 2001, ApJ, 552, 459

\bibitem[Heger \etal(2003)]{Heger03} Heger, A., Fryer, C. L., Woosley, S. E., Langer, N., \& Hartmann, D. H. 2003, ApJ, 591, 288

\bibitem[Johnson \& Bromm(2007)]{Johnson07} Johnson, J. L., \& Bromm, V. 2007, MNRAS, 374, 1557

\bibitem[Kassin \etal(2012)]{Kassin12} Kassin, S. A., Weiner, B. J., Faber, S. M., \etal 2012, ApJ 758, 106

\bibitem[Kawakatu \& Wada(2008)]{Kawakatu08} Kawakatu, N., \& Wada, K. 2008, ApJ, 681, 73  

%\bibitem[Komatsu \etal(2011)]{Komatsu11} Komatsu, E., \etal 2011, ApJS, 192, 18

\bibitem[Kormendy \& Ho(2013)]{Kormendy13} Kormendy, J., \& Ho, L. C. 2013, ARA\&A, 51, 511

\bibitem[Koushiappas \etal(2004)]{Koushiappas04} Koushiappas, S. M., Bullock, J. S., \& Dekel, A. 2004, MNRAS, 354, 292

\bibitem[Lodato \& Natarajan(2006)]{Lodato06} Lodato, G., \& Natarajan, P. 2006, MNRAS, 371, 1813

\bibitem[Loeb \& Rasio(1994)]{Loeb94} Loeb, A., \& Rasio, F. A. 1994, ApJ, 432, 52 

\bibitem[Madau \& Rees(2001)]{Madau01} Madau, P., \& Rees, M. J. 2001, ApJL, 551, L27

\bibitem[Mayer \etal(2010)]{Mayer10} Mayer, L., Kazantzidis, S., Escala, A., \& Callegari, S. 2010, Natur, 466, 1082

\bibitem[McGaugh(2005)]{McGaugh05} McGaugh, S. S. 2005, ApJ, 632, 859

\bibitem[McKinney \etal(2013)]{McKinney13} McKinney, J.C., Tchekhovskoy, A., Sadowski, A., \& Narayan, R. 2013, MNRAS, submitted (arXiv:1312.6127)  

\bibitem[Middleton \etal(2013)]{Middleton13} Middleton, M. J., Miller-Jones, J. C. A., Markoff, S., \etal 2013, Natur, 493, 187  

\bibitem[Milosavljevic \etal(2009)]{Milos09} Milosavljevic, M., Bromm, V., Couch, S. M., \& Oh, S. P. 2009, ApJ, 698, 766 

\bibitem[Mortlock \etal(2011)]{Mortlock11} Mortlock, D. J., Warren, S. J., Venemans, B. P., \etal 2011, Natur, 474, 616

\bibitem[Netzer \& Trakhtenbrot(2014)]{Netzer14} Netzer, H., \& Trakhtenbrot, B. 2014, MNRAS, 438, 672 

\bibitem[Novak(2013)]{Novak13} Novak, G. S. 2013, MNRAS, submitted (arXiv:1310.3833)
 
\bibitem[Page \etal(2013)]{Page13} Page, M. J., Simpson, C., Mortlock, D. J., \etal 2013, MNRAS, submitted (arXiv:1311.1686)
  
\bibitem[Pelupessy \etal(2007)]{Pelupessy07} Pelupessy, F. I., di Matteo, T., \& Ciardi, B. 2007, ApJ, 665, 107

\bibitem[Prieto \etal(2013)]{Prieto13} Prieto, J., Jimenez, R., \& Haiman, Z. 2013, MNRAS, 436, 2301 

\bibitem[Pringle(1981)]{Pringle81} Pringle, J. E. 1981, ARA\&A, 19, 137

\bibitem[Regan \& Haehnelt(2009)]{Regan09} Regan, J. A., \& Haehnelt, M. G. 2009, MNRAS, 396, 343 

\bibitem[Sadowski(2009)]{Sadowski09} Sadowski, A. 2009, ApJS, 183, 171

\bibitem[Sadowski \etal(2014)]{Sadowski14} Sadowski, A., Narayan, R., McKinney, J. C., \& Tchekhovskoy, A. 2014, MNRAS, 439, 503 

\bibitem[Shakura \& Sunyaev(1976)]{Shakura76} Shakura, N. I., \& Sunyaev, R. A. 1973, A\&A 24, 337

\bibitem[Shang \etal(2010)]{Shang10} Shang, C., Bryan, G. L., \& Haiman, Z. 2010, MNRAS 402, 1249

\bibitem[Shankar \etal(2010)]{Shankar10} Shankar, F., Crocce, M., Miralda-Escud\'e, J., Fosalba, P., \& Weinberg, D. H. 2010, ApJ, 718, 231
   
\bibitem[Shapiro(2005)]{Shapiro05} Shapiro, S. L. 2005, ApJ, 620, 59 

\bibitem[Soltan(1982)]{Soltan82} Soltan, A. 1982, MNRAS, 200, 115

\bibitem[Spaans \& Silk(2006)]{Spaans06} Spaans, M., \& Silk, J. 2006, ApJ, 652, 902  

\bibitem[Tanaka \& Haiman(2009)]{Tanaka09} Tanaka, T., \& Haiman, Z. 2009, ApJ, 696, 1798

\bibitem[Thorne(1974)]{Thorne74} Thorne, K. S. 1974, ApJ, 191, 507 

\bibitem[Volonteri \etal(2003)]{Volonteri03} Volonteri, M., Haardt, F., \& Madau, P. 2003, ApJ, 582, 559

\bibitem[Volonteri \& Rees(2005)]{VolonteriRees05} Volonteri, M., \& Rees, M. J. 2005, ApJ, 633, 624 

\bibitem[Volonteri \& Silk(2014)]{Volonteri14} Volonteri, M., \& Silk, J. 2014, ApJ, submitted (arXiv:1401.3513) 

\bibitem[Wada \& Norman(2002)]{Wada02} Wada, K., \& Norman, C. A. 2002, ApJL, 566, L21  

\bibitem[Wise \etal(2008)]{Wise08} Wise, J. H., Turk, M. J., \& Abel, T. 2008, ApJ, 682, 745 

\bibitem[Yoo \& Miralda-Escude(2004)]{Yoo04} Yoo, J., \& Miralda-Escude, J. 2004, ApJ, 614, L25 

\end{thebibliography}
\end{document}